\documentclass[iop, jcp, twocolumn, floatfix, superscriptaddress, reprint, 10pt]{revtex4-1}  
\usepackage{graphicx}
\usepackage[usenames]{color}
\usepackage{microtype}
\usepackage{amsmath}
\usepackage{amssymb}
\usepackage{xspace}
\usepackage[names]{xcolor}
\usepackage{footnote}
\usepackage{relsize}
\usepackage{hhline}

\newcommand{\be}{\begin{equation}}
\newcommand{\ee}{\end{equation}}

\newcommand{\NinitSF}{N}
\newcommand{\NredSF}{N'}
\newcommand{\ApproxX}{\Tilde{\mathbf{X}}}
\newcommand{\Xred}{\mathbf{X}'}
\newcommand{\D}{\ensuremath{\operatorname{d}}}
\usepackage{floatrow}

\begin{document}

\title{
Automatic Selection of Atomic Fingerprints and \\ Reference Configurations for Machine-Learning Potentials}

\author{Giulio Imbalzano}
\affiliation{Laboratory of Computational Science and Modeling, IMX, \'Ecole Polytechnique F\'ed\'erale de Lausanne, 1015 Lausanne, Switzerland}

\author{Andrea Anelli}
\affiliation{Laboratory of Computational Science and Modeling, IMX, \'Ecole Polytechnique F\'ed\'erale de Lausanne, 1015 Lausanne, Switzerland}

\author{Daniele Giofr\'e}
\affiliation{Laboratory of Computational Science and Modeling, IMX, \'Ecole Polytechnique F\'ed\'erale de Lausanne, 1015 Lausanne, Switzerland}

\author{Sinja Klees}
\affiliation{Lehrstuhl f\"ur Theoretische Chemie, Ruhr-Universit\"at Bochum, 44780 Bochum, Germany}

\author{J\"org Behler}
\affiliation{Universit\"at G\"ottingen, Institut f\"ur Physikalische Chemie, Theoretische Chemie, Tammannstr. 6, 37077 G\"ottingen, Germany}
\affiliation{Lehrstuhl f\"ur Theoretische Chemie, Ruhr-Universit\"at Bochum, 44780 Bochum, Germany}

\author{Michele Ceriotti}
\affiliation{Laboratory of Computational Science and Modeling, IMX, \'Ecole Polytechnique F\'ed\'erale de Lausanne, 1015 Lausanne, Switzerland}

\begin{abstract}
Machine learning of atomic-scale properties is revolutionizing molecular modelling, making it possible to evaluate inter-atomic potentials with first-principles accuracy, at a fraction of the costs. 
The accuracy, speed and reliability of machine-learning potentials, however, depends strongly on the way atomic configurations are represented, i.e. the choice of descriptors used as input for the machine learning method. The raw Cartesian coordinates are typically transformed in ``fingerprints'', or ``symmetry functions'', that are designed to encode, in addition to the structure, important properties of the potential-energy surface like its invariances with respect to rotation, translation and permutation of like atoms.
Here we discuss automatic protocols to select a %
number of fingerprints out of a large pool of candidates, based on the correlations that are intrinsic to the training data.
This procedure can greatly simplify the construction of  neural network potentials that strike the best balance between accuracy and computational efficiency, and has the potential to accelerate by orders of magnitude the evaluation of Gaussian Approximation Potentials based on the Smooth Overlap of Atomic Positions kernel. 
We present applications to the construction of  neural network potentials for water and for an Al-Mg-Si alloy, and to the prediction of the formation energies of small organic molecules using Gaussian process regression.
\end{abstract}

\maketitle

\section{Introduction}
In many applications of atomic-scale modelling, from electronic structure theory to molecular dynamics, the properties of molecular and condensed-phase structures are computed using as an input the Cartesian coordinates of the atoms, and the size of the periodic supercell. 
In many cases -- for instance when using density functional theory -- this description of the structures is perfectly suitable. 
For other types of potentials, like simple empirical force fields, Cartesian coordinates are usually discarded in favor of internal coordinates offering several advantages like a more intuitive connection between the structural description and the energy of a system.

In recent years, the widespread usage of machine-learning (ML) inspired techniques to recognize patterns in structures~\cite{rost+94prot,ball2010mach,chen06bion,gasp+18jctc,hara+10jctc,carr09mmm}, to classify molecules and to predict their properties based on a few reference calculations~\cite{huan+15prb,rupp12prl,fabe16prl,behl16jcp,de17jchmi,spar16scrm}, has moved the choice of coordinates again into the center of attention due to several deficiencies of these conventional coordinates~\cite{behl-parr07prl}. 
In order to be efficient in the interpolation between reference structures, and to achieve some degree of transferability, the representation that is used as the input of ML algorithms should encode the physical features, and the mandatory symmetries of the problem, such as invariance with respect to rotations, translations, and permutations of identical atoms~\cite{behl-parr07prl,bart+13prb,glie+17prb,gris+18prl}. 
To satisfy these requirements, many descriptors have been introduced that are able to characterize, to a various degree, atomic-scale systems, identify their similarities and differences, and form the basis for effective statistical learning schemes of energies and other properties~\cite{zhu+16jcp,huan-vonl16jcp,fabe+17jctc}.
Some of these descriptors, e.g. the symmetry functions used in the Behler and Parrinello neural network (NN) scheme~\cite{behl-parr07prl,behl11jcp}, come in the form of families of functions, that depending on their  parameters describe different correlations between particles within the environment. Choosing the set of parameters that characterizes the possible configurations in an equally economic and thorough way is one of the crucial steps in the optimization of ML schemes. 
Other descriptors, such as the Smooth Overlap of Atomic Positions (SOAP)~\cite{bart+13prb}, provide a systematically-converging representation of the environments, but the fingerprint vector can contain tens of thousands of elements, which increases considerably the computational costs. 

In this work we are going to discuss how it is possible to select the most suitable fingerprints to describe a given system, providing the best balance between computational cost and accuracy of predictions. 
In Section~\ref{sec:methods} we will briefly summarize the definition of the fingerprints we use as examples, and the methods we propose to select the best candidates to describe a given system and perform statistical learning of its properties.
In Section~\ref{sec:applications} we will present examples of applications to NN potentials for water and an Al alloy, and to the use of SOAP to generate a Gaussian Approximation Potential to predict molecular energetics. 
We then present our conclusions.

\section{Fingerprints for Machine Learning Potentials \label{sec:methods}}

\subsection{Behler-Parrinello Symmetry Functions}

The symmetry functions (SF) that are used in Behler-Parrinello neural network (BPNN) potentials have been introduced and described in great detail elsewhere~\cite{behl11jcp}. 
They have been used successfully as inputs for atomic feed-forward neural networks providing the atomic energy contributions to the system's total energy, and allowed to reproduce the stability~\cite{behl-parr07prl,khal+10prb,soss+12prb,eshe+12prl,artr-behl12prb,kapi+16jcp2,chen+16jpcl} or infrared spectra~\cite{gast+17cs} of different classes of materials and molecules.
Here we just summarize briefly the form and parameterization of two important families of Behler-Parrinello SFs which we will use in our examples, leaving the details of the fingerprints and the NN formulation to the many reviews on the subject~\cite{behl11jcp,behl11pccp,jose+12jcp,behler2017first}.

All the symmetry functions describe the correlations between atoms in a neighborhood of a central atom with index $i$.
The first functional form, called $G_2$ following the convention used in previous works~\cite{behl11jcp,behl2015ijqc,behler2017first}, provides information about pair correlations between the atoms
\begin{equation}
    G_2^i = \sum_j e^{-\eta (R_{ij} - R_s)^2} \cdot f_c (R_{ij}), \label{eq:g2}
\end{equation}
where the parameters $\eta$ and $R_s$ control the width and the position of the Gaussian with respect to the central atom and $ f_c (R_{ij})$ is a cutoff function that ensures that the symmetry function smoothly decreases to 0 in value and slope at a fixed cutoff $r_c$. The sum is over all neighboring atoms being closer than $r_c$.
The second type of symmetry functions, called $G_3$, provides information about angular correlations, and has the form
\begin{multline}
    G_3^i = 2^{1-\zeta} \sum_j \sum_{k \neq j } (1 + \lambda \cdot \cos \theta_{ijk})^{\zeta} \cdot \\
    e^{-\eta (R_{ij}^2 + R_{ik}^2 + R_{jk}^2)} \cdot f_c(R_{ij}) f_c(R_{ik}) f_c(R_{jk}), \label{eq:g3}
\end{multline}
where $\zeta$, $\eta$, and $\lambda$ are the three parameters that determine the shape of this type of symmetry function. The indices $j$ and $k$ run over all the atoms in the neighbourhood of the tagged atom $i$. The cutoff function that we have used has the form
\begin{equation}
    f_c(R_{ij}) = \left\{\begin{matrix}
\textup{tanh}^{3}\left [ 1 - \frac{ R_{ij} }{r_\text{c}} \right ] & \textup{for} & R_{ij} \leq  r_\text{c}\\ 
0.0 & \textup{for} & R_{ij} > r_\text{c}
\end{matrix}\right..
\end{equation}
The order of magnitude of the value of a SF depends trivially on its spatial extent and on the concentration of the species involved. We decided to normalize the value of each symmetry functions based on the value it would take if it were computed for a uniform ideal gas, so as to eliminate this trivial dependence and to treat them on more equal grounds. To do so, we first evaluate the integrals 
\begin{equation}
\begin{split}
I_2=& 4 \pi \rho_A \int \D r_A r_A^2 G_2(r_A)\\
I_3 = & 8\pi^2 \rho_A\rho_B \int \D r_A \D r_B \D \theta r_A^2 r_B^2 \sin\theta \ G_3(r_A, r_B, \theta), \\
\end{split} \label{eq:sf-integrals}
\end{equation}
where $\rho_A$ and $\rho_B$ are the average densities for the element corresponding to the first and second neighbor element considered in the evaluation of $G_2$ and $G_3$.
We then scale the symmetry functions by the square root of the integrals~\eqref{eq:sf-integrals} so as to guarantee that the variance of the values would be constant in a uniform gas. 

\begin{figure}[htbp]
  \centering
  \includegraphics[width=0.98\columnwidth]{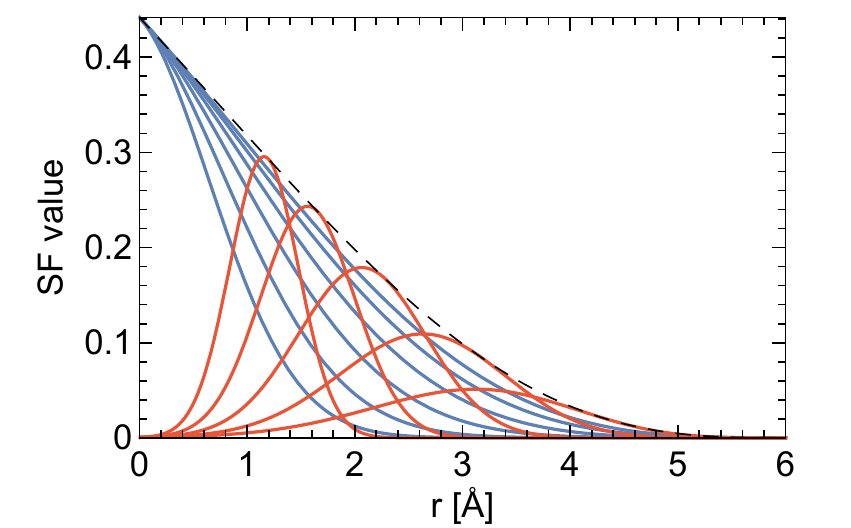}
  \caption{Examples of radial symmetry functions generated using $\NinitSF = 5$ and $r_\text{c} = 6\; \text{\AA}$. The blue curves are the symmetry functions centered in the origin ($R_s=0$) and $\eta$ varying as in Eq.~\ref{eqn:eta_central}, while the red ones have their center shifted using $R_s$ as described in Eq.~\ref{eqn:shifted_center} and $\eta$ is described by Eq.~\ref{eqn:eta_shifted}. The black dashed curve is the cutoff function for $r_\text{c} = 6\; \text{\AA}$.}
  \label{fig:symfun}
\end{figure}

Given that the method we propose is based on the sparsification of a large set of these SF fingerprints, a first preparatory step involves the determination of a thorough yet manageable pool of candidate SFs. 
The generation is done spanning over all of the meaningful sets of parameters, using simple heuristic rules to represent most of the possible correlations within the cutoff distance.
We generate two separate sets of radial symmetry functions, $G_2$. The first group is centered on the reference atom (i.e. $R_s = 0$) and the width varies as
\begin{equation} \label{eqn:eta_central}
    \eta_m = \left( \frac{n^{m/n}}{r_\text{c}} \right)^2,
\end{equation}
where $n$ is the number of intervals in which we have chosen to divide the space and $m= \{0,1,...,n\}$. The second group is centered along the path between the central atom and its neighbours, at increasing distances following
\begin{equation} \label{eqn:shifted_center}
    R_{s,m} = \frac{r_\text{c}}{n^{m/n}},
\end{equation}
while the Gaussian widths are chosen as
\begin{equation} \label{eqn:eta_shifted}
    \eta_{s,m} = \frac{1}{(R_{s,n-m}-R_{s,n-m-1})^2}
\end{equation}
in order to have narrow Gaussians close to the central atom and wider ones as the distances increases.
This effectively creates a finer grid closer to the central atom, where small variations in the position have a larger effect on the potential (see Fig.~\ref{fig:symfun}).

The $G_3$ symmetry functions were generated with a similar procedure, choosing values for  $\eta$ according to Eq.~\ref{eqn:eta_central}, setting $\lambda$ to both values $\{-1,1\}$ that were originally proposed and choosing a few values of $\zeta$ on a logarithmic scale. For instance, in the examples below we use $\{1,4,16\}$.

By increasing the cutoff radius and the number $\NinitSF$ of symmetry functions that are generated, one can make the description of the environment more and more complete. This comes however at the expense of greater computational costs, since a large number of SFs would then have to be generated at each potential evaluation. 
Less obviously, using too many, strongly correlated symmetry functions could lead to overfitting and difficulties in the regression process. We will discuss in Section~\ref{sec:selection} how to identify a small subset that conveys the essential structural information. 

\subsection{Smooth Overlap of Atomic Positions \label{SOAP}}

The Smooth Overlap of Atomic Positions (SOAP) framework has been introduced in Ref.~\cite{bart+13prb}, and has been used together with Gaussian Process Regression in many applications including metals, semiconductors, molecular crystals and small organic molecules~\cite{bart+10prl,deri-csan17prb,de+16pccp,bart+13prb2,szla+14prb,bart+17sa,musi+18cs}.
It is based on a representation of each local environment in terms of a smooth atomic probability amplitude $\psi_{\mathcal{X},\alpha}(\mathbf{r})$, which is constructed as a sum of Gaussians with a given variance $\sigma^2$, centered on each atom of element $\alpha$ within a given cutoff $r_\text{c}$ of the central atom $\mathbf{x}_0$.

\begin{equation}
    \psi_{\mathcal{X},\alpha}(\textbf{r}) = \sum_{i \in \mathcal{X}_\alpha} \textup{exp}\left(-\frac{\left[\textbf{r}-(\mathbf{x}_i -\mathbf{x}_0 )\right]^2}{2\sigma^2}\right).
\end{equation}

The amplitude $\psi_{\mathcal{X},\alpha}(\mathbf{r})$ constitutes in itself an infinite-dimensional fingerprint $\in \mathcal{L}^2$, invariant to translations and permutations of atomic labels within each element group. 
It is not however invariant to rotations, and the SOAP framework addresses this shortcoming by defining a kernel based on the average over all possible relative rotations  $\hat{R}$ of the overlap between the densities of two environments
\begin{equation}
    \Tilde{k}(\mathcal{X},\mathcal{X}') = \int d\hat{R}\left|\int\sum_\alpha \psi_{\mathcal{X},\alpha}(\textbf{r}) \psi_{\mathcal{X}',\alpha}(\hat{R}\textbf{r})\right|^n.\label{eq:soap-integral}
\end{equation}
Here, we use the symbol $\mathcal{X}$ to indicate them in an abstract manner.
The exponent $n$ must be chosen to be greater than one to preserve information on the relative angular orientation of atoms in the local neighbourhood. The kernel can be expanded to consider the overlap between different atomic species~\cite{de+16pccp}, which has been shown to increase the performance of the kernel for regression tasks~\cite{bart+17sa}, and to tensorial properties~\cite{gris+18prl}. 
Here, we restrict ourselves to the simpler formulation in which one considers only scalar properties, and the overlap between densities associated with the same element.
A very appealing feature of the framework is that, when one takes $n=2$, this integral can be written explicitly as the scalar product between the power spectra associated with the two environments
\begin{equation}
    \Tilde{k}(\mathcal{X},\mathcal{X}') = \sum_{\alpha\beta}\textbf{p}_{\alpha\beta}(\mathcal{X}) \cdot \textbf{p}_{\alpha\beta}(\mathcal{X}').
    \label{eq:soap-scalar}
\end{equation}
The vector $\textbf{p}_{\alpha\beta}(\mathcal{X})$ is computed based on the expansion coefficients for $\psi_{\mathcal{X},\alpha}$ and $\psi_{\mathcal{X},\beta}$ in a basis of radial functions and spherical harmonics, and expresses the information relative to the correlations between species $\alpha$ and $\beta$ in the local environment.
If the expansion contains $n_\text{max}$ radial functions, and maximum angular momentum channel $l_\text{max}$, $\textbf{p}_{\alpha\beta}$ contains $n_\text{max}^2 l_\text{max}$ elements, each of which constitutes a rotationally-invariant fingerprint for describing $\mathcal{X}$. 

The scalar-product form~\eqref{eq:soap-scalar} for the SOAP kernel converges systematically to the overlap integral~\eqref{eq:soap-integral} as the radial and angular cutoffs are increased. This comes however at considerable computational cost, since tens of thousands of power spectrum elements have to be computed and processed. 
It would therefore be useful to select a sub-set of the power spectrum elements that capture the essential information for a given system, even at the cost of loosening the formal connection between SOAP fingerprints and the integral kernel.

\section{Fingerprint Selection} \label{sec:selection}

Despite being based on very different premises, both sets of fingerprints discussed in the previous Section can lead to an arbitrarily high-dimensional feature space. 
We can now turn to the discussion of how one can select a $\NredSF$-dimensional subset of the initial fingerprints, that captures the essential features of the atomic environments and can be used with little or no performance loss as the basis of a statistical regression scheme to predict the properties of a given set of materials.
Symmetry functions can be selected, in a more or less automatic fashion, by evaluating empirically the accuracy of a ML model based on a trial set of SF. For instance, genetic algorithms have been recently proposed as a method to generate an optimal selection~\cite{gastegger2018jcp}, similar to what had been done in the past to select an optimal set of reference structures~\cite{Nick2016}. 
Here we focus on unsupervised approaches that rely only on knowledge of the geometries of the reference structures, without using information on energy and forces, nor on the performance of the ML model that results from a given choice of input features.

The first approach we will discuss here is based on a relatively simple idea: given a set of $M$ structures $\left\{ A_i\right\}$  that are representative of the system one wants to study, and a large number $N$ of fingerprints $\left\{\Phi_j\right\}$, one can build the $M\times N$ matrix $\mathbf{X}$ such that $X_{ij} = \Phi_j(A_i)$.
The most effective fingerprints can then be chosen by using standard linear algebra techniques to approximate $\mathbf{X}$. 
The ML schemes we discuss in this work are based on a decomposition of the energy of the system in local contributions, each of which is associated with an atom-centered spherical environment with cutoff radius $r_\text{c}$. 
Unless otherwise specified, we will consider these environments, rather than the entire structure, as the core of our discussion. The elements of $\mathbf{X}$ refer to the fingerprints defining these environments, which we consider, in order to simplify the notation, without explicit reference to the structure they are part of.
A given set of features can be used in a variety of regression schemes to predict atomic-scale properties, ranging from linear fits to Gaussian Process regression~\cite{bart+10prl} (GPR) and neural networks~\cite{behl-parr07prl,smith2017ani} (NN). 
It has been shown that in many cases the quality of the input representation plays a much more important role than the regression algorithms in determining the accuracy of predictions~\cite{fabe+17jctc}.
Here we will consider examples of both GPR and NN schemes, applied to the two paradigmatic families of atomic fingerprints described above.

We aim to find the optimal $M\times \NredSF$ feature matrix $\Xred$, where $\NredSF\ll \NinitSF$, that still provides a satisfactory representation of the space while reducing the computational load of the ML scheme. 
This is essentially a dimensionality reduction problem, that can be interpreted in terms of the construction of a low-rank approximation $\ApproxX$ of the feature matrix. 
Most of the dimensionality techniques available for this task, such as singular value decomposition (SVD), generate new features that are a linear combination of the initial set and cannot be used for our current purpose, as they would still require the evaluation of all the $\NinitSF$ features and, only as a second step, project them onto a lower-dimensional space.
We have therefore considered methods that strive to obtain a low-rank approximation of the feature matrix or its associated covariance using only rows and columns of $\mathbf{X}$. We discuss in particular three approaches, namely CUR decomposition, farthest point sampling (FPS) and a Pearson correlation (PC)-based method.

\subsection{CUR Decomposition}

CUR decomposition~\cite{maho-drin09pnas} is a feature selection method that has been developed to deal with data where the information provided by the singular vectors cannot be properly interpreted, such as gene expression data. 
In analogy with the low-rank approximation obtained with a singular value decomposition, one writes
\begin{equation}\label{eqn:CURformula}
    \mathbf{X} \approx \ApproxX = \mathbf{C}\, \mathbf{U}\, \mathbf{R}
\end{equation}
where $\mathbf{C}$ and $\mathbf{R}$ are actual rows and columns of the original matrix. The objective is still to find the best low-rank approximation to $\mathbf{X}$, but in this case only actual elements of the matrix are used, which implies that $\ApproxX$ can be obtained without having to compute all $\NinitSF$ fingerprints. 

We discuss in particular the procedure for selecting a reduced number of columns (i.e. fingerprints), but the method can also be used to reduce the number of rows (i.e. reference structures)~\cite{bart-csan15ijqc}.
Each column $c$ of the initial feature matrix is given an ``importance score'' calculated as
\begin{equation}
    \pi_c = \sum_{j=1}^k (\nu_c^{(j)})^2,
\label{eq:cur-score}
\end{equation}
where $\nu_c^{(j)}$ is the $c$-th coordinate of the $j$-th right singular vector, and $k$ is the number of features that have yet to be selected and runs from $\NredSF$ to $1$. We also found that a very effective selection can be obtained by using a fixed number of singular vectors $k=1$ at each iteration in the procedure (CUR$(k=1)$). Not only this makes the method numerically more stable and significantly faster, but it makes the selection independent on the target number of symmetry functions, so that one can effectively perform a single selection with a large $N'$, obtaining a list of SF that is sorted from the most important to the least important.
The importance score can also be weighted by a factor if one wants to prioritize the selection of a certain type of features, e.g. if the cost of evaluating different fingerprints varies greatly, and one would rather take several ``cheap'' fingerprints than a single ``expensive'' one.

Most CUR schemes employ a probabilistic criterion for feature selection, to guarantee e.g. that if several nearly-identical features are present, any of them will have approximately the same probability of being selected. 
To obtain a deterministic selection, we pick at each step the column with the highest score, and avoid selecting multiple nearly-identical features with an orthogonalization procedure. 
After having selected the $l$-th column with the highest importance score, every remaining column in $\mathbf{X}$ is orthogonalized relative to it
\begin{equation}
    X_j \leftarrow X_j - X_l\; ( X_l \cdot X_j )/\left|X_l\right|^2.
\end{equation}
The SVD is then re-computed based on the orthogonalized matrix, and the column weights are re-evaluated. The procedure is iterated until all $\NredSF$ features have been chosen to build the $\mathbf{C}$ matrix, that corresponds to the reduced feature matrix $\Xred$.
Since in this application we are only interested in reducing the number of fingerprints, $\mathbf{R}=\mathbf{X}$, and we can compute  $\mathbf{U} = \mathbf{C}^+ \mathbf{X} \mathbf{X}^+$, where $\mathbf{A}^+$ indicates the pseudoinverse. One can then compute the accuracy of the approximation as
\begin{equation}
\epsilon = \left\| \mathbf{X} - \mathbf{C}\mathbf{U}\mathbf{R} \right\|_F/\left\| \mathbf{X}\right\|_F
\end{equation}
The total number of features to be selected, can either be fixed a priori, or increased until $\epsilon$ becomes smaller than a prescribed threshold.

\subsection{Farthest Point Sampling}
Alternatively, one can select the features using a farthest-point sampling (FPS) approach. This is analogous to the strategy with which one can select uniformly-spaced reference points (see e.g.~\cite{ceri+13jctc}), but here we apply it to the \emph{columns} of $\mathbf{X}$, so as to select fingerprints that are as diverse as possible for the data set being investigated. 
In a FPS scheme, successive points are chosen so as to maximize the Euclidean distance between them.  After arbitrarily selecting the first fingerprint, each subsequent one is chosen as
\begin{equation}
    k = \operatorname{argmax}(\operatorname{min}_j|X_k-X_j|),
\end{equation}
where $j$ refers to all of the features that have already been selected. The procedure is repeated until all $\NredSF$ features have been chosen. 

\subsection{Pearson Correlation Method}

Finally, we propose a third method for the selection of the features, based on the Pearson correlation. In this PC method, features from the pool of candidate functions are determined one after the other in such a way to minimize the correlation of the feature values of the available data set.
The Pearson correlation $P_{i,j}$ of two features $\Phi_i$ and $\Phi_j$ for atoms of the same element is given by
\begin{eqnarray}
P_{i,j}=\frac{\sum_{k=1} \left( \Phi_{i}(\mathcal{X}_k)-\bar{\Phi}_i\right)\cdot \left(\Phi_{j}(\mathcal{X}_k)-\bar{\Phi}_j \right)}{N_{\mathrm{atom}} \cdot \sigma(\Phi_i)\cdot \sigma(\Phi_j)} \quad.
\end{eqnarray}
Here, $\Phi_{i}(\mathcal{X}_k)$ is the value of $\Phi_i$ for the k-th atom in the whole data set, $\bar{\Phi}_i$ is the arithmetic mean of the values of $\Phi_i$ over all atoms of the respective element, and $\sigma(\Phi_i)$ is the standard deviation.

Specifically, we start by picking an arbitrary feature from the pool as the first, which in the case of SF has been chosen as the radial function $G_2$ with $R_s=0$, the smallest available $\eta$ value and the largest cutoff $r_c$ to start from the SF with the largest spatial extent. Then the second function to be added to the set is the one with the smallest correlation to this function. The following functions are then selected one after the other such that the average correlation to the functions already included in the set is smallest. This procedure is followed for each element in the system separately.
The accuracy of each method in selecting the most relevant symmetry functions can be tested empirically by comparing the accuracy of the regression model built using a given SF selection. Alternatively, one can also compute the expression~(\ref{eq:cur-score}) to assess the error in approximating the full feature matrix.  
Finally, it should also be stressed that the PC, FPS and CUR($k=1$) methods are all greedy selection strategies that effectively generate a sorted list of features, so that one can always start with a relatively large selection and remove the least important features until the best compromise between accuracy and evaluation cost is achieved.

\subsection{Global Fingerprints and Train Set Selection}

As mentioned before, one could use CUR, FPS or PC methods to sparsify the train set, that is to reduce the number of reference structures rather than the number of fingerprints. This can be useful to reduce the cost of evaluating a ridge regression model, or to minimize the number of property evaluations that need to be performed in order to train the model~\cite{bart-csan15ijqc,de+16pccp,bart+17sa}. 
In order to do so, it is useful to construct a set of fingerprints associated with the whole structure, rather than with individual atomic environments. A straightforward definition of a ``global'' fingerprint associated with a structure $A$, $\bar{\Phi}(A)$ is the average of all the local fingerprints for the environments that compose the structure $A$, i.e.
\begin{equation}
    \bar{\Phi}_j(A) = \sum_{\mathcal{X}_k \in A} \Phi_j(\mathcal{X}_k)/N_{\text{at}}(A).
\label{eq:fp-global}
\end{equation}

In the case of Behler-Parrinello symmetry functions, that are defined separately for each chemical species, we consider that the global fingerprint is composed by concatenating sections corresponding to each element. 
In other terms, one can see this as a sparse representation for a larger fingerprint vector that is padded with zeros in all sections but the relevant one, even though in a practical NN implementation one only computes symmetry functions associated with the identity of the central atom.
The fingerprint vector for the entire structure can then be built according to \eqref{eq:fp-global}, summing these zero-padded vectors over all atoms in the structure.

\section{Applications \label{sec:applications}}

\subsection{A Potential for Liquid Water}

As a first example, we consider the case of liquid water. 
For this system we can compare our approaches to the SFs of a previously published NN potential that has been built out of carefully-chosen fingerprint functions.
This potential, that has been trained on a DFT reference data set~\cite{mora+16pnas}, and that has been applied to study a variety of properties of liquid water,
provides a remarkably concise description of water environments, consisting of only 32 $G_2$ and 25 $G_3$ functions. Using the same or similar symmetry functions, also alternative parameterizations have provided excellent results for water~\cite{kapi+16jcp2,chen+16jpcl}, electrolytes~\cite{P4670} and even solid-liquid interfaces~\cite{nata+15pccp,P4988}.

In order to identify automatically suitable sets of fingerprints for water, we started by taking the same data set that was used in Ref.~\cite{kapi+16jcp2}, and selected by FPS a set of 1000 structures that we use for symmetry function selection and training. 
We generated an initial pool of 768 SF combining three sets of $G_2$ functions obtained following the protocol discussed in Section~\ref{sec:methods}, with $N=8$ and cutoffs $r_c=4,8,12$ bohr, 
and two sets of $G_3$ SF generated with $N=8$ -- one with $r_c=4$ bohr and $\zeta=1,2,4,8,16$ and one $r_c=8$ bohr and $\zeta=1,2,4$.
Final results are not sensitive to these choices, that we only made to have intermediate files of manageable size. We removed duplicate SFs, those with a length scale smaller than 0.75~\AA{}, and normalized the values by the square root of their uniform-gas average value. 
We weighted the importance scores~\eqref{eq:cur-score} by a factor proportional to $\rho_A \rho_B r_c^3$ for $G_2$ functions between atoms $A$ and $B$ and $\rho_A \rho_B \rho_C r_c^6$ for $G_3$ functions between atoms $A$, $B$ and $C$, to reflect the cost associated with evaluating them. We note that these importance scores do not enter the functional form of the SFs finally used in the fit.

For assessing the performance of the optimized SFs, we selected SF sets containing $\NredSF=$ 16, 32, and 64 symmetry functions for each element using a CUR, FPS and a PC procedure. 
Additionally, for comparison we include a ``default'' symmetry function set in our benchmark, which we frequently use for first preliminary potentials. For a binary system like water this default set contains 6 $G_2$ functions for each element pair with parameters $\eta$ chosen such that the turning points of the terms in the summation in Eq.~\ref{eq:g2} are equidistantly arranged between the minimum interatomic distance and the function with maximum spatial extension ($\eta=0$). $R_s$ is set to zero and $r_c=12$ bohr. For the angular functions $G_3$ (Eq.~\ref{eq:g3}) we use for each possible element combination the parameter sets $\zeta =1,2,4,16$ along with $\lambda=\pm 1$, $\eta=0$ and $r_c=12$ bohr. Therefore, the default set contains each 36 SFs for the oxygen and hydrogen atoms. Finally, also the SFs of Ref.~\cite{mora+16pnas} have been tested with our reference data set.

For each set of symmetry functions we trained 4 NN potentials based on atomic NNs with two hidden layers and 20 neurons per hidden layer using the RuNNer code~\cite{runner}, with random initial weights and a 3:1 random split of train:test points using the same 1000 FPS subset.
Table~\ref{table:SFWater-CUR} reports the average test error for energy and forces obtained using the CUR, FPS and PC SFs sets as well as of the ``default'' set and the SF set of Ref.~\cite{mora+16pnas}. The table also shows the CUR approximation errors for O and H fingerprints for each number of symmetry functions, and the execution time per MD step for a simulation with 216 water molecules ran using the LAMMPS RuNNer plugin~\cite{plim95jcp,mora+16pnas} on a single Intel Xeon 2.60GHz core.

\begin{table}[btph]
\setlength\tabcolsep{3 pt}
\resizebox{0.98\columnwidth}{!}{%
\centering
\begin{tabular}{ccccc}
\hline \hline
$N'_\text{O},N'_\text{H}$  &
$\begin{array}{c}
\epsilon_\text{O},\epsilon_\text{H}\\
\times 10^{-4}
\end{array}$
& $\begin{array}{c}
\text{RMSE}(E)\\
\text{[meV/at.]}
\end{array}$ & 
$\begin{array}{c}
\text{RMSE}(f)\\
\text{[eV/\AA]}
\end{array}$
& $\begin{array}{c}\text{Runtime}\\
\text{[s/step]}
\end{array}$ \\ \hline
\multicolumn{5}{l}{CUR selection}\\\hline
16,16       & 51,63     & 1.55     & 0.147   &    0.35  \\ 
32,32       & 2.5,6.2   & 1.18     & 0.126   &    0.43  \\ 
64,64       & 0.1,0.3   & 0.99     & 0.114   &    0.52  \\ 
\hline
\multicolumn{5}{l}{CUR$_{k=1}$ selection}\\\hline
16,16       &  51,63    &  1.49    & 0.145   &    0.35  \\ 
32,32       &  2.6,7.6  &  1.23    & 0.123   &    0.42  \\ 
64,64       &  0.1,0.3  &  1.02    & 0.113   &    0.52  \\ 
\hline
\multicolumn{5}{l}{FPS selection}\\\hline
16,16       & 56,132    & 3.89     & 0.251   &    0.34  \\ 
32,32       & 7.1,12    & 1.62     & 0.150   &    0.40  \\ 
64,64       & 0.3,0.9   & 1.19     & 0.128   &    0.51  \\ 
\hline
\multicolumn{5}{l}{PC selection}\\\hline
16,16       & 248,341   & 2.81     & 0.232   &    0.37  \\ 
32,32       & 93,85     & 1.43     & 0.146   &    0.42  \\ 
64,64       & 0.3,34    & 1.11     & 0.123   &    0.53  \\ 
\hline
\multicolumn{5}{l}{Default SF set}\\\hline
36,36       &           & 1.62     & 0.238   &    0.85  \\ 
\hline \multicolumn{5}{l}{SFs of Ref. \citenum{mora+16pnas}}\\\hline
30,27       & -         & 0.98     & 0.115   &    0.69  \\\hline\hline
\end{tabular}
\caption{The table reports, for different numbers of SF selected from a pool of 768 candidates using different strategies, the error in the approximation of the feature matrix, and the RMSE for energies and forces from a test set, averaged over four NNs trained starting from different random weights. The spread between results of the 4 independent training runs for each choice of SF is of the order of 2-4\%. Results from the SF used in Ref.~\citenum{mora+16pnas} and of a ``default set'' are also shown for comparison. \label{table:SFWater-CUR}}
}
\end{table}

All the different strategies to automatically select fingerprints show that it is possible to progressively improve the test set accuracy by making the selection more inclusive. 
CUR gives by far the best performance, both in terms of error in approximating $\mathbf{X}$ and in terms of the energy and force test RMSE, followed by PC and then FPS. 
Interestingly, PC gives marginally better results than FPS on force/energy predictions, even though it shows considerably worse approximation error for the feature matrix. 
All the automatic selection protocols perform better than the ``default'' SF set, dramatically so in the case of CUR.

However, manual optimization of symmetry functions, taking into account the physical parameters of the system, and the actual accuracy of the training, seems to provide an advantage. 
The selection from Ref.~\cite{mora+16pnas} achieves with only 57 SF the same accuracy as a CUR selection of 128. The automatic selection, however, requires a lower computational effort, since the estimated cost of evaluating a SF is taken into account when generating the selection.
It would be possible to further improve the performance of the automatic selection by considering also the correlations between the SF values and the target property, such as energies or forces -- so as to select the descriptors that are not only structurally uncorrelated, but also strongly coupled to the stability of the system. 

\subsection{A Potential for Aluminum Alloys}

Water is a two-component system, but its molecular nature means that the number of possible environments is affected less dramatically by the number of species.
A NN potential for Al-Si-Mg alloys has been recently demonstrated~\cite{koba+17prm}, that instead deals with a ternary system, where all of the interactions among the different species and defects must be accounted for to obtain accurate predictions across the full range of relevant compositions.
The presence of multiple interactions at different length scales makes the manual selection of SF a particularly cumbersome task. %
In the previous work~\cite{koba+17prm}, the problem was circumvented by restricting the SF pool to the 2-body $G_2$ components, 
making it possible to obtain a systematic - if not optimal - selection. The automatic selection procedure we introduce in this work makes it much easier to automatically determine an efficient feature set that includes both $G_2$ and $G_3$ SFs, which makes it possible to take into account the angular dependence of the atomic interactions explicitly.

The reference data set we used as a starting point is composed of the 10551 structures used by Kobayashi et al.~\cite{koba+17prm}, supplemented by 609 structures of $\beta''$-phase precipitates and interfaces that have been generated in a previous DFT study of the alloy~\cite{giof+17am}. Given that many of the resulting 11160 structures are taken from short MD runs and are highly correlated, we selected 2000 structures with FPS, that have been used both for the selection of SF and the training/testing procedure.
This sparser selection leads to a larger absolute magnitude of the fit error, but does not affect the quality of the fit, while making the optimization procedure faster and more stable. 

The initial generation of SF is done similarly to the case of water. Six sets of $G_2$ SF have been generated using $N = 4,12$ and $r_c = 8,16,20$ bohr, and two sets of $G_3$ SF have been generated using $N = 8$ - one with $r_c = 8$ bohr and $\zeta=1,2,4,8,16$ and the other with $r_c = 12$ bohr and $\zeta=1,2,4$. 
Duplicate SF have been eliminated, together with those that had a width smaller than 1.06~\AA{} for the radial ones and smaller than 1.32~\AA{} for the angular ones. The same weighting described for water has been used here when selecting the SF.
The details of the fingerprints can be inferred from the input files provided in the SI, and the performance of the resulting NN potentials can be seen in Table~\ref{table:al6xxx}.
The test set RMSE decreases systematically as the number of selected SF increases, up to 64 SFs per species. We also compared the results with those obtained with the SF selection from  Ref.~\citenum{koba+17prm}.
To ensure a fair comparison we re-optimized and tested the potential using the RuNNer~\cite{runner} software and the same FPS selection we discuss above. 
Already at $N'=96$ (32 SFs per species) the automatic selection that includes 3-body SFs leads to a better test set error than the systematic selection of 120 $G_2$ SFs.

\begin{table}[btph]
\setlength\tabcolsep{3 pt}
{%
\centering
\begin{tabular}{cccc}
\hline \hline
$N'_\text{Al},N'_\text{Mg},N'_\text{Si}$  &
$\begin{array}{c}
\epsilon_\text{Al},\epsilon_\text{Mg},\epsilon_\text{Si}\\
\times 10^4
\end{array}$
& $\begin{array}{c}
\text{RMSE}(E)\\
\text{[meV/at.]}
\end{array}$ & 
$\begin{array}{c}
\text{RMSE}(f)\\
\text{[eV/\AA]}
\end{array}$ \\ \hline
\multicolumn{4}{l}{CUR selection}\\\hline
16,16,16       & 79,99,101     & 16.22     & 0.084   \\ 
32,32,32       & 7.9,14,10   & 4.08     & 0.052   \\ 
64,64,64       & 0.9,1.3,0.8   & 2.47     & 0.022   \\ 
\hline
\multicolumn{4}{l}{SFs of Ref. \citenum{koba+17prm}}\\\hline
40,40,40       & -         & 9.2     & 0.069  \\\hline\hline
\end{tabular}
\caption{The table reports, for different numbers of SF, the error in the approximation of the feature matrix, and the RMSE for energies and forces from a test set. Results from the SF used in Ref.~\cite{koba+17prm} are also shown for comparison.\label{table:al6xxx}}
}
\end{table}

While the test set RMSE is a good measure of the quality of a potential, it is important to also verify the stability of the NN when computing a property for which configurations had not been explicitly included in the train set. 
As an example of the behavior of the different potentials, that is very relevant for the potential application of this NN in the description of the early stages of precipitation in Al-6xxx alloys~\cite{giof+17am}, we computed the configuration energy along the minimum energy pathways for the vacancy-assisted migration of Al, Si, Mg atoms in a matrix of 256 Al atoms.  
Atomic configurations along the pathway between the minimum energy states were obtained by linear interpolation, and by local optimization using the nudged elastic band (NEB) method~\cite{henk-jons99jcp} with the climbing image algorithm~\cite{henk+00jcp} as implemented in Quantum ESPRESSO~\cite{gian+09jpcm}. 
The details of the DFT calculations were the same as described in Refs.~\cite{giof+17am,koba+17prm}. 7 images have been used for Mg and 13 have been used for Al and Si, and lead to relaxed vacancy migration barriers that are consistent with previous DFT calculations~\cite{mantina2009}.
Keeping the configurations fixed, we computed the energy along the migration barrier for both the linear transition path between the initial and final configurations and the corresponding relaxed positions. 

\begin{figure}
    \centering
    \includegraphics[width=0.98\columnwidth]{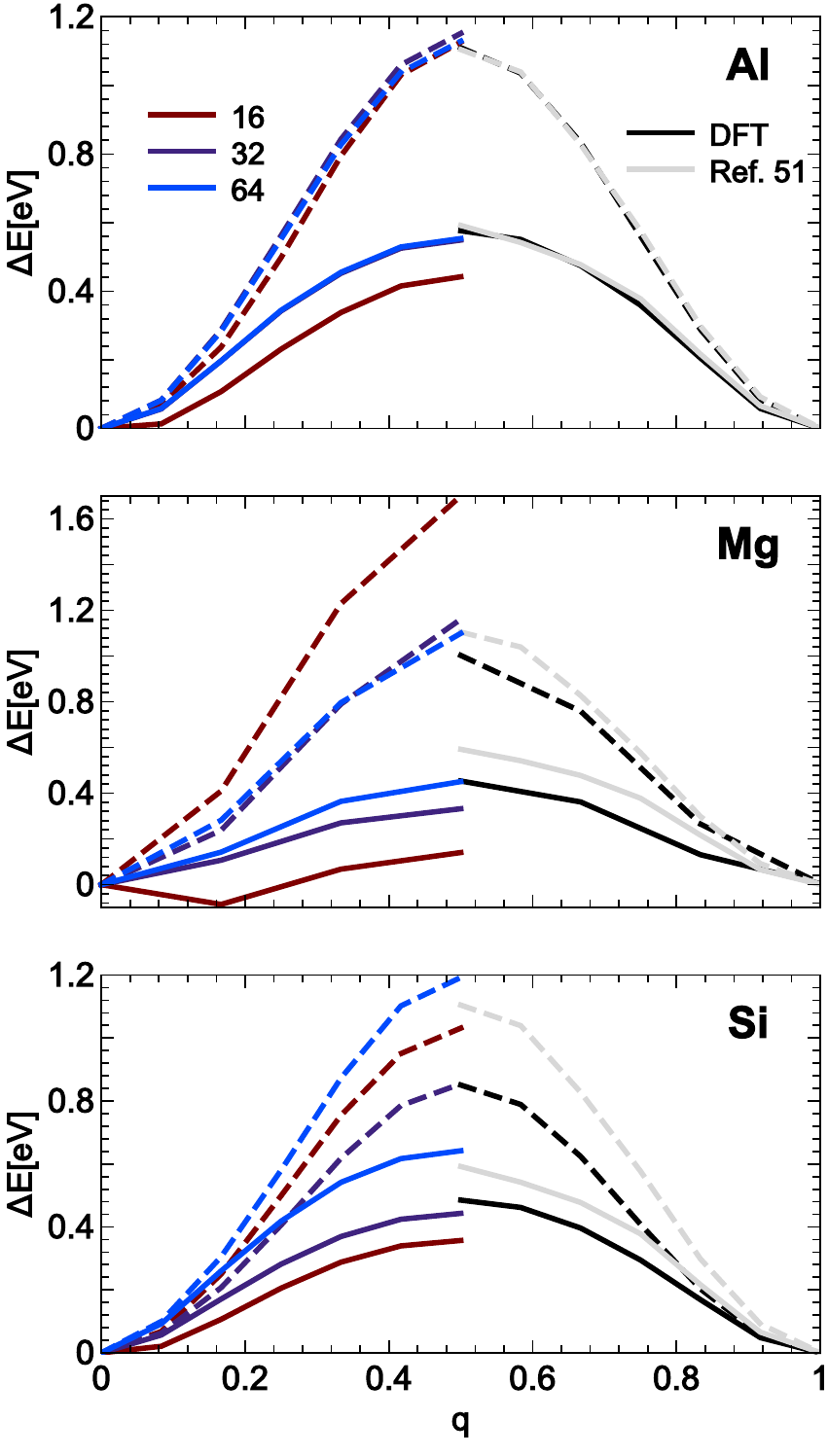}
    \caption{The energy barrier for the vacancy-assisted migration of Al, Mg, and Si using an increasing number of symmetry functions are presented on the left, compared to DFT and the choice of SF from Ref.~\citenum{koba+17prm}, presented on the right. Dashed lines correspond to the unrelaxed configurations, solid lines to the minimum energy pathway. The energies are shown as a difference from the minimum energy structure.}
    \label{fig:vacancy-neb}
\end{figure}

As shown in Figure~\ref{fig:vacancy-neb}, there is a considerable improvement in the quality of the fit when going from 16 to 32 SFs per species, whereas the improvement is less dramatic when using a larger number of SFs, and actually in the case of the vacancy-assisted diffusion of Si the 64-SF NN performs worse than the 32-SF NN. This observation underscores the fact that refining the SF selection does systematically improve the accuracy in the interpolative regime, as probed by cross-validation, but not necessarily to a systematic improvement in the extrapolative regime. 
For all of the vacancy-assisted diffusion processes we considered, however, NN potentials reproduce the correct qualitative behavior. Excluding the case with 16 SF per element, which is clearly insufficient for this system, the error in the relaxed barrier is below 0.1 eV, which is comparable to the typical DFT error. Automatic SF selections that include 3-body terms perform better than the $G_2$-only choice of Ref.~\citenum{koba+17prm}, that nevertheless predicts diffusion barriers with a remarkably small error.

\subsection{Learning Molecular Energies}

To provide a very different example of the application of dimensionality reduction strategies to sparsify the feature matrix, we turn to the case of SOAP fingerprints, and to the GPR of atomization energies for a molecular data set composed of 7211 small organic molecules, containing up to 7 heavy atoms (N, C, O, Cl, S)~\cite{mont+13njp}.
As we discussed in Section~\ref{sec:methods}, the SOAP framework provides a very systematic method to describe a chemical environment, but can easily lead to thousands of descriptors. 
In this case, which involves 6 chemical species, and for which we used an environment cutoff of $r_c=3.0~\text{\AA}$, $n_\text{max}=9$ and $l_\text{max}=9$, one has to deal with a total of $\NinitSF=14852$ rotationally-invariant symmetry functions. 
This huge number of features is in stark contrast with the handful of symmetry functions that are used in the BPNN scheme to generate accurate interatomic potentials.
It is reasonable to speculate that a small fraction of the initial features could also provide a satisfactory description of the chemical environments, and therefore an accurate prediction of properties. 
To test this idea, we have applied the same framework we discussed for the BP symmetry functions to the power spectrum $\textbf{p}_{\alpha\beta}(\mathcal{X})$.

There is however an important difference compared to the previous case. For BPNN the feature vectors are fed through a neural network, and are subject to a linear transformation before being fed to the first layer of non-linear activation functions, whose coefficients are optimized together with all the other parameters of the NN. 
SOAP fingerprints are typically used to compute a kernel for Gaussian process regression, that in its simplest form corresponds to the scalar product between features, without an optimization step to determine the most effective linear combination of the inputs.
For this reason, in order to reduce the size of the input vectors without compromising the regression accuracy, it is necessary to introduce an additional ingredient. 
The original kernel is calculated as $\mathbf{K} = \mathbf{X}\mathbf{X}^T$, whereas now we intend to compute it using the approximate form of $\mathbf{X}$, i.e. we intend to find $\Tilde{\mathbf{K}} = \ApproxX\ApproxX^T$, where $\ApproxX$ is shown in eq \ref{eqn:CURformula}.

As previously explained in section \ref{sec:selection}, given that we only aim to reduce the number of features, $\mathbf{U}\mathbf{R}=\mathbf{C}^+\mathbf{X}$. The approximate kernel can then be written as
\begin{equation}
    \Tilde{\mathbf{K}} = \mathbf{C}\mathbf{C}^+\mathbf{X}\mathbf{X}^T(\mathbf{C}^+)^T\mathbf{C}^T.
\end{equation}
Computing the approximate kernel also involves the $\NredSF\times\NredSF$ matrix $\mathbf{W} = \mathbf{C}^+\mathbf{X}\mathbf{X}^T(\mathbf{C}^+)^T$. 
Since this matrix is symmetric and positive-definite, it can be decomposed as $\mathbf{W}=\mathbf{A}\mathbf{A}^T$. 
Finally, we see that the kernel can be written in terms of scalar products of the reduced-dimensionality features, provided we define $\Xred=\mathbf{C}\mathbf{A}$, since
\begin{equation}
    \Tilde{\mathbf{K}} = (\mathbf{C}\mathbf{A})(\mathbf{C}\mathbf{A})^T \quad .
\end{equation}
Therefore, after using the previously described schemes to select features from $\mathbf{X}$, we also have to compute the $\mathbf{A}$ matrix in order to scale adequately the selected features. 
It should be noted that the matrix $\mathbf{A}$, although computed only once during the fingerprint selection stage, must be stored and applied to the selected components of the power spectrum when performing training or predictions.

\begin{figure}[hpbt] 
  \centering 
  \includegraphics[width=0.98\columnwidth]{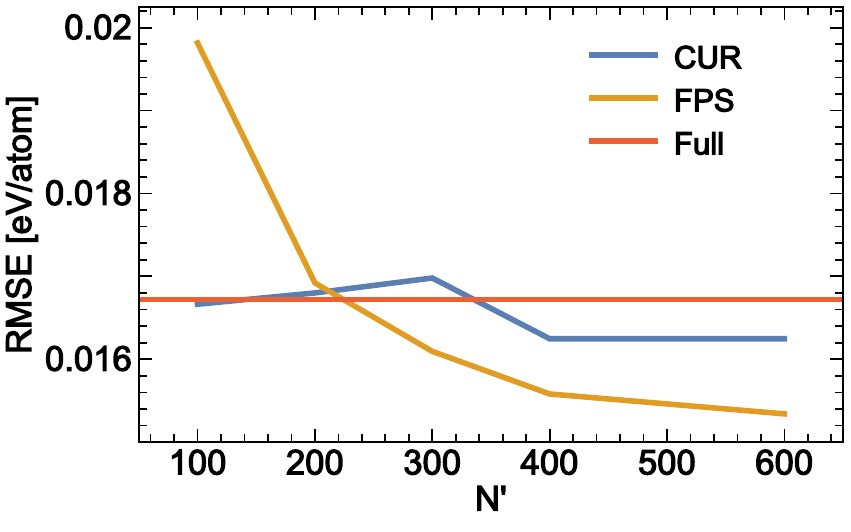}
  \caption{The RMSE of the GPR for 1442 randomly chosen structures in the test set, with a varying number of elements of the power spectrum, chosen for both CUR and FPS, compared to the result of the GPR with the full power spectrum. The training set is composed by 500 FPS structures.\label{fig:feature-selection}}
\end{figure}

Let us now turn to discuss the performance of different feature selection strategies for the prediction of the atomization energies on the QM7b data set.
All the results we present are tested using the same set, composed of 1442 randomly selected structures (which correspond to roughly 20\% of the full QM7b data).
From the overall training set, containing 5769 structures, we select 500 structures with a FPS strategy that we use to construct the initial feature matrix. 
We then apply both the CUR and the FPS methods to perform feature selection, and use the reduced dimensionality set of descriptors to train a GPR model on the same 500 FPS structures. Figure~\ref{fig:feature-selection} shows the RMSE in the prediction of the atomization energies of the test-set structures. 
It is remarkable to see that using only 100 CUR-selected elements of the power spectrum it is possible to match the prediction accuracy obtained with the original kernel based on more than 14,000 features. 
Interestingly, increasing the number of features to 400 leads to \emph{lower} test error, suggesting that for this small training set the use of a smaller set of fingerprints helps to combat overfitting. FPS selection also performs remarkably well, and at $\NredSF=400$ it yields a test RMSE which is 5\%{} lower than the baseline SOAP result. 

\begin{figure}[hpbt] 
  \centering
  \includegraphics[width=0.98\columnwidth]{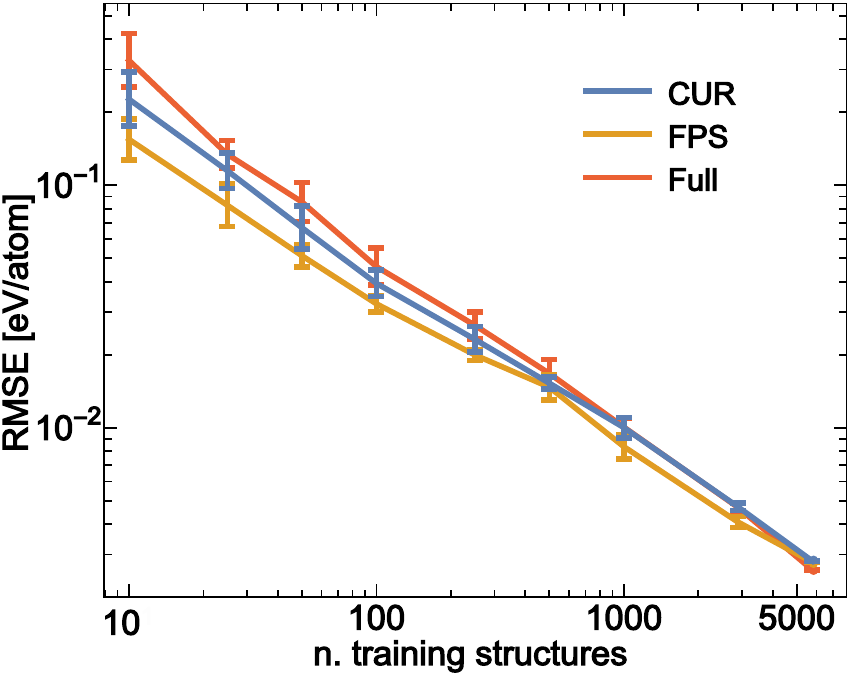}
  \caption{Learning curves for the QM7b atomization energies~\cite{mont+13njp}, when using the full SOAP power spectrum, 400 features selected with FPS, and 400 selected with CUR. The results shown for each train set size are the average and standard deviation from 10 different models trained on random selections extracted from the overall training set  \label{fig:learning-curve-SOAP}}
\end{figure}

The question is of course whether this reduced-dimensionality description is sufficient to further improve the prediction accuracy, when more structures are used for training. 
As seen in the learning curves in Fig.~\ref{fig:learning-curve-SOAP}, using 400 features is enough to obtain errors that are comparable to the reference value, or even lower. 
It is only when considering the full 5769 structures in the training set that the baseline kernel reaches a marginally better accuracy than the reduced-dimensionality model, that discards as much as 97\%{} of the elements of the SOAP power spectrum.

\section{Conclusions}

In this article we have discussed methods that can be used to select the most important features out of a large initial pool of candidates. %
This unsupervised scheme is based on preserving the features that retain the most information, and are identified using relatively standard linear algebra methods, which makes the approach fully automatic and transferable to different systems and families of descriptors.
Given that our strategy determines a low-dimensional description that requires the evaluation of only a small number of features, it can also reduce dramatically the cost of the property prediction.

The examples we use demonstrate the applicability of this scheme to machine-learning scenarios, that differ widely in terms of descriptors, regression scheme and nature of the atomistic data.  
The first case we discuss is that of Behler-Parrinello symmetry functions, which have been used as descriptors of atomic environments and inputs for neural network interatomic potentials. In our example we consider the case of a potential for condensed phases of water, for which a very successful reference set of symmetry functions exists. 
The choice of the parameters for the symmetry functions
is currently tackled either by selecting the best symmetry functions based on an understanding of the physics of the system at hand and experience~\cite{behl2015ijqc}, or using a systematic generation of parameters~\cite{smith2017ani}.
By selecting automatically a small number of BP symmetry functions out of a large pool of automatically-generated descriptors we could achieve the same prediction error of the reference physically-motivated choice of symmetry functions, while reducing the cost of evaluating them.
The advantage of using an automatic procedure to select SFs is even more apparent when considering the case of the ternary alloy of Al, Si and Mg. 
NN potentials based on CUR-selected SFs outperform a previous potential, both in terms of test RMSE and in terms of the accuracy of predicting vacancy-assisted atom migration barriers.  

We then considered as a different application the sparsification of the very large feature matrix that corresponds to the SOAP power spectrum, and used as benchmark the atomization energies of a set of small organic molecules. 
We show that by retaining only 3\%{} of the power spectrum components one can match or outperform a Gaussian process regression model that uses the entire kernel. 
These results suggest that the SOAP framework, that provides a complete, systematically-converging representation of chemical environments, can be used at a much reduced cost by selecting a small set of components that provide a sufficient amount of chemical information. 

Our scheme, that could be easily adapted to other classes of atomic and molecular descriptors, and that could also be used to select features from multiple classes of fingerprints,  simplifies and/or accelerates greatly the optimization and the use of machine-learning models of atomic-scale properties. 
On a more fundamental level, it suggests that the success of regression schemes as surrogate models for interatomic potentials can be understood in terms of a relatively low dimensionality of the manifold that describes energetically-accessible molecular motifs. 
Analyzing the nature of this low-dimensional manifold to reveal a more intuitive understanding of structure-property relations~\cite{musi+18cs}, quantifying the sensitivity of the manifold to the reference data set, and if necessary proposing a strategy to actively adapt the feature set to the accumulation of new structures during a simulation are all promising directions to further extend this line or research.

\begin{acknowledgments}
MC and AA were supported by the European Research Council under the European Union's Horizon 2020 research and innovation programme (grant agreement no. 677013-HBMAP).
DG and MC acknowledge support for this work by an Industrial Research Grant funded by Constellium. 
GI acknowledges funding from the Fondazione Zegna. SK was supported by the Cluster of Excellence RESOLV (EXC 1069) funded by
the Deutsche Forschungsgemeinschaft, JB is grateful for a DFG Heisenberg professorship (Be3264/11-2).
\end{acknowledgments}

\end{document}